\documentclass[11pt]{article}
\usepackage{amsfonts}
\usepackage{amssymb}
\usepackage{amssymb,amsfonts,amsmath,amsthm,cite,color}
\usepackage{dsfont}
\usepackage{epsfig}
\usepackage{mathrsfs}
\newtheorem{thm}{Theorem}[section]

\newtheorem{lem}[thm]{Lemma}
\newtheorem{prop}[thm]{Proposition}
\newtheorem{defn}[thm]{Definition}

\newtheorem{ex}[thm]{Example}

\makeatletter \@addtoreset{equation}{section}

\def\pf{\noindent {\it Proof.\ }}
\def\qed{\hfill \rule{4pt}{7pt}}

\newcommand{\bQ} { {\mathbb{Q}}}

\def\Disp{{\rm Disp}}

\def\res{{\rm res}}

\begin{document}

\begin{center}

 {\Large \bf An Algorithm for Deciding the Summability of\\[5pt]
  Bivariate Rational Functions
}
\end{center}
\begin{center}
{  Qing-Hu Hou}$^{1}$ and {Rong-Hua Wang}$^{2}$

   $^1$Center for Applied Mathematics\\
   Tianjin University, Tianjin 300072, P.R. China \\
   hou@nankai.edu.cn \\[10pt]

   $^2$Center for Combinatorics, LPMC-TJKLC\\
   Nankai University, Tianjin 300071, P.R. China\\
   wangwang@mail.nankai.edu.cn

\end{center}

\vskip 6mm \noindent {\bf Abstract.} Let $\Delta_x f(x,y)=f(x+1,y)-f(x,y)$ and $\Delta_y f(x,y)=f(x,y+1)-f(x,y)$ be the difference operators with respect to $x$ and $y$. A rational function $f(x,y)$ is called summable if there exist rational functions $g(x,y)$ and $h(x,y)$ such that $f(x,y)=\Delta_x g(x,y) + \Delta_y h(x,y)$. Recently, Chen and Singer presented a method for deciding whether a rational function is summable. To implement their method in the sense of algorithms, we need to solve two problems. The first is to determine the shift equivalence of two bivariate polynomials.
We solve this problem by presenting an algorithm for computing the dispersion sets of any two bivariate polynomials. The second is to solve a univariate difference equation in an algebraically closed field. By considering the irreducible factorization of the denominator of $f(x,y)$ in a general field, we present a new criterion which requires only finding a rational solution of a bivariate difference equation. This goal can be achieved by
deriving a universal denominator of the rational solutions and a degree bound on the numerator. Combining these two algorithms, we can decide the summability of
a bivariate rational function.


\section{Introduction}
In 1978, Gosper~\cite{Gosper1978} presented the celebrated algorithm which solves the problem of determining whether a given hypergeometric term is
equal to the difference of another hypergeometric term. Based on Gosper's algorithm, Zeilberger~\cite{Zeilberger1990c, Zeilberger1991}
gave a fast algorithm for proving terminating hypergeometric identities. Zeilberger's method was further extended to the multivariate case by Wilf and Zeilberger himself
in~\cite{Wilf1992}.  Paule~\cite{Paule1995b} gave an interpretation of Gosper's algorithm in terms of the greatest factorial factorizations.
Chen, Paule and Saad~\cite{ChenPauleSaad2008} derived an easy understanding version of Gosper's algorithm by considering the
convergence of the greatest common divisors of two polynomial sequences.

Other approaches to the summability of rational functions were given by Abramov \cite{Abramov1975, Abramov1995, Abramov1995b}, Pirastu and Strehl \cite{Pirastu1995b}, Ash and Catoiu \cite{AshCatoiu2005}. The key idea of these methods is to rewrite a rational function $\alpha$ as $\alpha=\Delta(\beta)+\gamma$,
where $\Delta$ is the difference operator, $\beta$ and $\gamma$ are rational functions such that the denominator of $\gamma$ is shift-free. Then $\alpha$ is summable if and only if $\gamma$ is zero.

Passing from univariate to multivariate, Zeilberger's algorithm have been discussed by
Zeilberger himself \cite{MohammedZeilberger2005, Apagodu2006}, Koutschan \cite{Koutschan2010},
Schneider~\cite{Schneider2005}, Chen et. al.~\cite{ChenHouMu2006}. These algorithms are useful in practice. However, they did not provide a complete answer to
the summability problem of bivariate hypergeometric terms. Only very recently, Chen and Singer~\cite{ChenSinger2014} presented criteria for deciding the summability of \emph{bivariate rational functions}. When applying their criteria, one will encounter two problems. The first one is how to determine whether two bivariate polynomials are shift equivalent. The second one is how to solve univariate difference equations in algebraically closed fields. The main aim of the present paper is to overcome these problems and to give an algorithm for deciding the summability of bivariate rational functions. We remark that the general question considered in this paper was raised by
Andrews and Paule in~\cite{AndrewsPaule1993}.

For the first problem, we show that the dispersion set of two bivariate polynomials is computable. Then two polynomials are shift equivalent if and only if the dispersion set is not empty. For the second problem, we present a variation of the criteria by considering the irreducible factorization in a general field instead of an algebraically closed field. To apply the new criteria, we need only to find rational solutions of a bivariate difference equation. By a discussion similar to Gosper's algorithm, we derive a universal denominator of the rational solutions. We further derive a degree bound on the numerator of the rational solutions and thus obtain an algorithm for the new criterion. Combining these two algorithms, we finally obtain an algorithm for deciding the summability of bivariate rational functions.

The paper is organized as follows. In Section \ref{sec:shifteq}, we give an algorithm for computing the dispersion set of two bivariate polynomials. In Section \ref{sec:criterion}, we first reduce the summability of a general rational function to that of a rational function whose denominator is a power of an irreducible polynomial. Then we present a criterion on the summability of this special kind of rational functions. This criterion reduces the summability problem to the problem of finding rational solutions of a bivariate difference equation. In Section \ref{sec:solve}, we give an algorithm for solving the difference equation.

Throughout the paper, we take $\mathbb{Q}$, the field of rational numbers, as the ground field.
It should be mentioned that the discussions work also for other fields, such as the extension
field $\mathbb{Q}(\alpha_1, \ldots, \alpha_r)$ where $\alpha_1,\ldots,\alpha_r$ are either algebraic or transcendental
over $\mathbb{Q}$.

We follow the notations used in~\cite{ChenSinger2014}.
Let~$f(x,y) \in \mathbb{Q}(x,y)$ be a bivariate rational function.
The shift operators $\sigma_x$ and $\sigma_y$ are given by
\[
\sigma_xf(x,y)=f(x+1,y)\quad  {\rm and} \quad \sigma_yf(x,y)=f(x,y+1).
\]
A function $f\in \mathbb{Q}(x,y)$ is said to be \emph{$(\sigma_x,\sigma_y)$\hbox{-}summable} if there exist two rational functions $g,h\in \mathbb{Q}(x,y)$ such that
\[f=\sigma_x g-g +\sigma_y h -h .\]

\section{Dispersion set and shift equivalence}\label{sec:shifteq}
Let $\mathbb{Z}$ denote the set of integers.
Recall that given two univariate polynomials, say $f(x)$ and $g(x)$, their \emph{dispersion set} is defined by
\[
\Disp_x(f,g) = \{n \in \mathbb{Z} \mid   f(x) = g(x+n) \}.
\]
It is known that unless $f$ and $g$ are the same constant polynomial, the dispersion set $\Disp_x(f,g)$ is finite and is computable.
For the algorithm, see~\cite[page 79]{PWZbook1996}. We can extend this concept to the bivariate case.

\begin{defn}\label{equivalent}
Let $f,g$ be two bivariate polynomials in $\mathbb{Q}[x,y]$ and $\sigma_x,\sigma_y$ be the shift operators. The \emph{dispersion set} of $f$ and $g$ is defined by
\[
\Disp (f,g) = \{(m,n) \in \mathbb{Z}^2 \mid f = \sigma_x^m \sigma_y^n g \}.
\]
If~$\Disp (f,g)$ is not empty, we say $f$ and $g$ are \emph{shift equivalent}.

In particular, when
$f=\sigma_x^mg$ (resp.\ $f=\sigma_y^ng$), we say $f,g$ in the same
$\sigma_x$-\emph{orbit}~(resp.\ $\sigma_y$-\emph{orbit}), denoted by $f\sim_xg$ and $f\sim_yg$ respectively.
\end{defn}

We remark that testing shift equivalence over fields have been considered by Grigoriev and Karpinski \cite{GrigorievKarpinski1993, Grigoriev1997, Grigoriev1996}. More precisely, they gave algorithms to find shifts $(\alpha_1,\ldots,\alpha_r) \in F^r$ such that
\[
f(x_1+\alpha_1, \ldots, x_r+\alpha_r)=g(x_1,\ldots,x_r),
\]
where $F$ is a field and $f,g \in F[x_1,\ldots,x_r]$. Instead of considering shifts over a field, we focus on integer shifts, i.e., $m,n \in \mathbb{Z}$.

In the univariate case, the dispersion set of any two polynomials is computable. The following theorem shows that the dispersion set is also computable in the bivariate case.

\begin{thm}\label{(m,n)-equ-judge}
Let $f, g \in \mathbb{Q}[x,y]$ be two polynomials. Then we can determine the dispersion set~$\Disp (f,g)$.
\end{thm}
\pf Since the shift operators $\sigma_x$ and $\sigma_y$ preserve the degree, we get that $\Disp(f,g) = \emptyset$ unless $\deg_x f = \deg_x g$.

When $f=0$ or $\deg_x(f)=0 $, the computation of $\Disp(f,g)$
reduces to the univariate case. More precisely, we have
\[
\Disp(f,g) = \mathbb{Z} \times \Disp_y(f,g).
\]

Now assume that $\deg_x f = d >0$ and write~$f, g$ as
\[
 f =\sum_{k=0}^{d}a_k(y)x^k, \quad g =\sum_{k=0}^{d}b_k(y)x^k.
\]
Suppose that $(m,n) \in \Disp(f,g)$. By comparing the leading coefficient with respect to $x$, we see that $n$ falls in the dispersion set
\[
{\cal N} = \Disp_y(a_d(y), b_d(y)).
\]
If ${\cal N}$ is a finite set, we then have
\[
\Disp(f,g) = \bigcup_{n_0 \in {\cal N}} \Disp_x(f(x,y), g(x,y+n_0)) \times \{n_0\}.
\]
Otherwise, we may assume $a_d(y)=b_d(y)=c$, where $c$ is a non-zero constant.
By comparing the second leading coefficient with respect to $x$, we see that
\begin{equation}\label{eq-m}
  a_{d-1}(y)=d\cdot c \cdot m+b_{d-1}(y+n).
\end{equation}
According to the degree of $a_{d-1}(y)$ in variable $y$, there are three cases.

\noindent \emph{Case $1$.} $\deg a_{d-1}(y)>1$. Then $\Disp(f,g) =\emptyset$ unless the leading term of $a_{d-1}(y)$ and that of $b_{d-1}(y)$ coincide. Assume
\[
    a_{d-1}(y) = \sum_{j=0}^h p_j y^j \quad \mbox{and} \quad b_{d-1}(y) = \sum_{j=0}^h q_j y^j.
\]
By comparing the coefficients of $y^{h-1}$ in the expansions of $a_{d-1}(y)$ and $b_{d-1}(y+n)$, we see that $n$ is uniquely determined by
\begin{equation}\label{eq-n0}
 h q_h n + q_{h-1}= p_{h-1}.
\end{equation}
Suppose that $n_0$ is an integer solution of \eqref{eq-n0}. We then have
\[
 \Disp(f,g) = \Disp_x(f(x,y), g(x,y+n_0)) \times \{n_0\}.
\]

\noindent \emph{Case $2$.} $\deg a_{d-1}(y) = 1$. We also have $\Disp(f,g)=\emptyset$ unless the leading term of $a_{d-1}(y)$ and that of $b_{d-1}(y)$ coincide. Assume
\[
a_{d-1}(y) = p_1 y + p_0 \quad \mbox{and} \quad b_{d-1}(y) = p_1 y + q_0.
\]
Then \eqref{eq-m} leads to
\begin{equation}\label{EQ:diophantine}
    (d \cdot c)\cdot m + p_1 \cdot  n = p_0 - q_0,
\end{equation}
which is a linear Diophantine equation in unknowns $m,n$. Either there is no solution, or the solutions are of the form
\[
m=u t + v, \quad  \mbox{and} \quad n=u' t + v',
\]
where $u,v,u',v'$ are explicit integers and $t$ runs over $\mathbb{Z}$. Now by setting all coefficients of $x,y$ in the expansion of $f(x,y) - g(x+ut+v,y+u't+v')$ to be zeros,
we arrive at a system of polynomial equations in $t$. The set of integer solutions of each equation is computable (see, for example \cite[page 79]{PWZbook1996}). The final dispersion set of $f$ and $g$ is the intersection of these solution sets.

\noindent \emph{Case $3$.} $\deg a_{d-1}(y) = 0$ or $a_{d-1}(y)=0$. If $\deg_y b_{d-1}(y) > 0$, we then have $\Disp(f,g)=\emptyset$. Otherwise, $m$ is uniquely determined by \eqref{eq-m}. Suppose $m_0$ is an integer solution of~\eqref{eq-m}, we have
    \[
    \Disp(f,g) = \{m_0\} \times \Disp_y(f(x,y), g(x+m_0,y)).
    \]

This completes the proof. \qed

Based on the proof as above, we can describe an algorithm for computing the dispersion set of two polynomials in~$\mathbb{Q}[x,y]$.

\noindent \emph{Algorithm DispSet}

\noindent \emph{Input:} Two polynomials $f=\sum_{k=0}^{d_1} a_k(y) x^k$ and $g=\sum_{k=0}^{d_2} b_k(y) x^k$.

\noindent \emph{Output:} The dispersion set~$\Disp(f, g)$.

\begin{enumerate}
\item
If $d_1 \neq d_2$, return $\emptyset$. Else set $d=d_1=d_2$.

\item
If $d \leq 0$, return the set $\mathbb{Z} \times \Disp_y(f,g)$.
Else continue the following steps.

\item
If $\deg a_{d}(y)>0$, compute ${\cal N } = \{n \mid a_d(y) = b_d(y+n) \}$ and for each $n_0 \in \cal N$, compute the set $S_{n_0}$ of integers $m$ such that $ f = \sigma_x^m \sigma_y^{n_0} g$.
Return the set
\[
\bigcup_{n_0 \in {\cal N}} S_{n_0} \times \{n_0\}.
\]
Else set $a_d(y)=c$ and continue the following steps.

\item
If $\deg_y a_{d-1}(y) > 1$,
compute the unique $n_0$ according to \eqref{eq-n0}. If $n_0$ is an integer, then return $\Disp_x(f(x,y), g(x,y+n_0)) \times \{n_0\}$. Else return $\emptyset$.

\item
If $\deg_y a_{d-1}(y)=1$. If the leading terms of $a_{d-1}(y)$ and $b_{d-1}(y)$ are different, then return $\emptyset$.
Else solve the linear Diophantine equation~\eqref{EQ:diophantine}. Suppose that the solutions are of the form
\[
m=u t + v \quad \mbox{and} \quad n=u' t + v'.
\]
Substituting $m$ by $ut+v$ and $n$ by $u't+v'$ in $f=\sigma_x^m \sigma_y^n g$ and comparing the coefficients of each power of $x$ and $y$ to get a system of polynomial equations in $t$. Return all integer solutions if there are. Else return $\emptyset$.
\item
If $\deg_y a_{d-1}(y)=0$ or $a_{d-1}(y)=0$. If $\deg_y b_{d-1}(y) > 0$ then return $\emptyset$. Else compute the unique $m_0$ satisfying~\eqref{eq-m}. If $m_0$ is not an integer, then return $\emptyset$. Else return the set
\[
\{m_0\} \times \Disp_y(f(x,y), g(x+m_0,y)).
\]

\end{enumerate}

The following is an example which shows how to determine the shift equivalence of any two given bivariate polynomials.

\begin{ex}\label{EX:shift-equivalent}
Let
\[
f=2x^2+2xy+y^2+y+1\quad \mbox{and} \quad g=2x^2+2xy+y^2+2x+y+1.
\]
We try to determine whether $f$ and $g$ are shift equivalent according to the proof of Theorem~\ref{(m,n)-equ-judge}. Rewrite $f,g$ as
\[
f=2x^2+(2y)x+(y^2+y+1),\ and \quad g=2x^2+(2y+2)x+(y^2+y+1).
\]
It's easy to check that this meets Case $2$ in the proof. Thus $m,n$ satisfy the linear equation $2m+n=-1$ whose solutions are
\[
m=t\ \mbox{and} \ n=-2t-1, \quad t \in\mathbb{Z}.
\]
Now by setting all coefficients
of $x$, $y$ in the expansion of $f(x,y)-g(x+t, y-2t-1)$ to be zeros, we obtain an integer solution $t=-1$. It means that $f(x,y)=g(x-1,y+1)$ and thus $f,g$ are shift equivalent.
\end{ex}

\section{Summability criterion}\label{sec:criterion}
As stated in the introduction, one can decompose a univariate rational function $\alpha$ into the form $\alpha=\Delta \beta + \gamma$. The goal of this section is to introduce a bivariate variant of such additive decomposition
and thus reduce the bivariate summability problem of a general rational function to that of a rational function whose denominator is a power of an irreducible polynomial. We then present a criterion for the summability of this kind of special rational functions.

Let $f\in \bQ(x, y)$ be a bivariate rational function. Assume that
the irreducible factorization of the denominator $D(x,y)$ of $f(x,y)$ is
\[
D(x,y) = \prod_{i=1}^m d_i^{n_i}(x,y),
\]
where $d_i(x,y)$ are irreducible polynomials and $n_i$ are positive integers. Viewing $f$ as a rational function of $y$ over the field $\mathbb{Q}(x)$, we have the partial fraction decomposition
\begin{equation}\label{EQ:pfd}
f = P + \sum_{i=1}^m \sum_{j=1}^{n_i} \frac{a_{i, j}}{d_i^j},
\end{equation}
where~$P \in \bQ(x)[y]$, $a_{i, j}\in \bQ(x)[y]$ and $\deg_y(a_{i,j})<\deg_y(d_i)$. It is well known that the polynomial $P$ is the difference of a polynomial.

Now suppose that $d_i(x,y) = d_k(x+m,y+n)$ for some index $i \not= k$. Then we have
\[
\frac{a_{i,j}}{d_i^j} = \sigma_x(g)-g + \sigma_y(h)-h + \frac{\sigma_x^{-m} \sigma_y^{-n} (a_{i,j})}{d_k^j},
\]
where
\[
g = \begin{cases}
  {\sum\limits_{\ell=0}^{m-1}} \frac{\sigma_x^{\ell-m}(a_{i,j})}{\sigma_x^\ell \sigma_y^n(d_k^j)}, & \mbox{if $m \ge 0$}, \\[15pt]
  - \sum\limits_{\ell=0}^{-m-1} \frac{\sigma_x^\ell(a_{i,j})}{\sigma_x^{m+\ell} \sigma_y^n(d_k^j)}, & \mbox{if $m < 0$},
\end{cases}
\]
and
\[
h = \begin{cases}
  {\sum\limits_{\ell=0}^{n-1}} \frac{\sigma_y^{\ell-n} \sigma_x^{-m}(a_{i,j})}{\sigma_y^\ell (d_k^j)}, & \mbox{if $n \ge 0$}, \\[15pt]
  - \sum\limits_{\ell=0}^{-n-1} \frac{\sigma_y^\ell \sigma_x^{-m}(a_{i,j})}{\sigma_y^{n+\ell} (d_k^j)}, & \mbox{if $n < 0$}.
\end{cases}
\]
Repeating the above transformation, we arrive at the following decomposition.
\begin{lem}\label{LM:xy-reduce}
For a rational function~$f\in \bQ(x,y)$, we can decompose it into the form
\[ f=\Delta_x(g) + \Delta_y(h) +r ,\]
where $g, h \in \bQ(x,y)$ and $r$ is of the form
\begin{equation}\label{EQ:resform}
r =\sum_{i=1}^m\sum_{j=1}^{n_i}\frac{a_{i,j}(x,y)}{d_i^j(x,y)},
\end{equation}
with $a_{i,j}\in \bQ(x)[y]$, $\deg_y(a_{i,j})<\deg_y(d_i)$, $d_i\in \bQ[x,y]$ are irreducible polynomials,
and~$d_i$ and~$d_{i'}$ are not shift equivalent
for any $1\leq i\neq i' \leq m$.
\end{lem}

From Lemma~\ref{LM:xy-reduce}, we see that $f $ is $(\sigma_x,\sigma_y)$\hbox{-}summable
if and only if $r$  is $(\sigma_x,\sigma_y)$\hbox{-}summable.
The following lemma shows that the summability of $r$ is equivalent to the summability of each summand of $r$.

\begin{lem}\label{LEM:split}
Let $r \in \mathbb{Q}(x,y)$  be of the form \eqref{EQ:resform}. Then $r$ is $(\sigma_x,\sigma_y)$\hbox{-}summable  if and
only if $\frac{a_{i,j}(x,y)}{d_i^j(x,y)}$ is $(\sigma_x,\sigma_y)$\hbox{-}summable for all $1\leq i\leq m$ and $1 \leq j \le n_i$.
\end{lem}
\pf The sufficiency follows from the linearity of the difference operators $\Delta_x$ and $\Delta_y$. It suffices to prove the necessity. Assume that~$r$ is $(\sigma_x,\sigma_y)$-summable,
then there exist $g,h\in \mathbb{Q}(x,y)$ such that~$r=\sigma_x(g)-g+ \sigma_y(h)-h$. We can always decompose~$g, h$ as
\[
g =\frac{A_1 }{D_1 }+\frac{A_2 }{D_2 } \quad \hbox{ and }\quad
h =\frac{B_1 }{C_1 }+\frac{B_2 }{C_2 },
\]
where $A_i,B_i,C_i,D_i (i=1,2)$ are polynomials in $y$ over $\mathbb{Q}(x)$, $\deg_y(A_1)<\deg_y(D_1)$, $\deg_y(B_1)<\deg_y(C_1)$, $D_1$ (resp.\ $C_1$) contains only irreducible factors that are shift equivalent to $d_i$, while $D_2$ (resp.\ $C_2$) contains no such factors.
Let $r_i=
\sum_{j=1}^{n_i}\frac{a_{i,j}(x,y)}{d_i^j(x,y)}$. We then have
\begin{align*}
  r_i -\left(\sigma_x\frac{A_1 }{D_1 }-\frac{A_1 }{D_1 }
   +\sigma_y\frac{B_1 }{C_1 }-\frac{B_1 }{C_1 }\right)
= \sigma_x\frac{A_2 }{D_2 }-\frac{A_2 }{D_2 }
    +\sigma_y\frac{B_2 }{C_2 }
    -\frac{B_2 }{C_2 }-\sum_{j\neq i}r_j.
\end{align*}
Note that~$\sigma_x, \sigma_y$ preserve the $(\sigma_x, \sigma_y)$-equivalence.  Therefore, we have
\[
r_i =\sigma_x\frac{A_1 }{D_1 }-\frac{A_1 }{D_1 }
      +\sigma_y\frac{B_1 }{C_1 }-\frac{B_1 }{C_1 },
\]
which means $r_i$ is $(\sigma_x,\sigma_y)$-summable.

By the same observation as in~\cite[Page 330]{ChenSinger2014}, we see that~$\sigma_x$ and~$\sigma_y$ preserve
the multiplicities of the fractions~$a_{i,j}/d_i^j$.
This implies that $r_i$ is $(\sigma_x, \sigma_y)$-summable
if and only if each summand $a_{i,j}/d_i^j$ is $(\sigma_x, \sigma_y)$-summable. This concludes the proof.
\qed

Now we only need to study the summability problem of rational functions of the form $a/d^j$, where $d\in \bQ[x, y]$ is irreducible, ~$a\in \bQ(x)[y]$, and $\deg_y(a)<\deg_y(d)$. For this kind of rational functions, we have the following criterion for their summability.
\begin{thm}\label{theorem2}
Let $ f=\frac{a(x,y)}{d^j(x,y)}$, where $d(x,y)\in \mathbb{Q}[x,y]$ is an
irreducible polynomial, $a\in \mathbb{Q}(x)[y]$ is non-zero and  $\deg_y(a)<\deg_y(d)$. Then $ f $ is $(\sigma_x,\sigma_y)$-summable if and only if
\begin{itemize}
\item[{\rm (1)}]
there exist integers $t,\ell$ with $t \not= 0$ such that
\begin{equation}\label{c1}
  \sigma_x^{t}d(x,y)=\sigma_y^{\ell}d(x,y),
\end{equation}
\item[{\rm (2)}]
for the smallest positive integer $t$ such that \eqref{c1} holds, we have
\begin{equation}\label{c2}
a=\sigma_x^{t}\sigma_y^{-\ell}p-p,
\end{equation}
for some $p\in \mathbb{Q}(x)[y]$ with $\deg_y(p)<\deg_y(d)$.
\end{itemize}
\end{thm}

We can adapt the argument used in~\cite[Theorem 3.7]{ChenSinger2014} to complete the proof of Theorem~\ref{theorem2}. The details are elaborated in the appendix.

The criterion~\eqref{c1} can be tested by computing the dispersion set $\Disp(d,d)$. In the next section, we will give an algorithm for solving the equation~\eqref{c2}. Then combining Lemma~\ref{LM:xy-reduce}, Lemma~\ref{LEM:split} and Theorem~\ref{theorem2}, we will obtain an algorithm for determining whether a bivariate rational function is summable.

\section{Rational solutions of the difference equations}\label{sec:solve}
Let $d_0$ be a positive integer and $u$ be a polynomial in $y$ over $\mathbb{Q}(x)$ with $\deg_y(u)<d_0$.
In this section, we present a method of finding solutions $p \in \mathbb{Q}(x)[y]$ with $\deg_y(p)<d_0$ to the following difference equation
\begin{equation}\label{EQ:denom}
u = \sigma_x^m \sigma_y^{-n} p - p,
\end{equation}
where $m,n$ are given integers and $m>0$.

Noting that $\deg_y(p)<d_0$, we may assume
\[
p = p_0(x) + p_1(x)y + \cdots + p_{d_0-1}(x)y^{d_0-1}.
\]
Then comparing the coefficients of each power of $y$ on both sides of \eqref{EQ:denom}, we obtain a system of linear difference equations in $p_i(x)$. Abramov-Barkatou \cite{AbramovBarkatou1998} and Abramov-Khmelnov \cite{AbramovKhmelnov2012} presented algorithms for solving such systems.

We will rewrite $p$ as the ratio $c(x,y)/d(x)$ and estimate the denominator $d(x)$ directly. Then we give an upper bound on the $x$-degree of the numerator $c(x,y)$ and thus solve for $p$ by the method of undetermined coefficients.

Assume that $u=a(x,y)/b(x)$, where $a,b$ are polynomials in $x$ and $y$. We notice that one can give an estimation of $d(x)$ by using the convergence argument introduced by Chen, Paule and Saad \cite{ChenPauleSaad2008}. More precisely, we have
\[
d(x) \, \big| \, \gcd\big( b(x)b(x+m) \ldots, b(x-m)b(x-2m) \cdots \big).
\]
Note also that one can give an estimation by using an argument similar to \cite{HouMu2011}. Here we give another estimation based on Gosper representation \cite[page 80]{PWZbook1996}.

Rewrite Equation~\eqref{EQ:denom} as
\begin{equation}\label{a(x,y)2}
a(x,y)=\frac{b(x)}{b(x+m)}\sigma_x^m\sigma_y^{-n}(b(x)p(x,y))-b(x)p(x,y).
\end{equation}
Let
\begin{equation}\label{Gosper expansion}
\frac{b(x)}{b(x+m)}=\frac{A(x)}{B(x)}\frac{C(x+m)}{C(x)},
\end{equation}
be the Gosper representation. That is,
\begin{equation}\label{gcd}
{\rm gcd}(A(x),B(x+hm))=1, \quad \forall \ h\in \mathbb{N}.
\end{equation}
Then the denominator of $p$ can be given by the following theorem.

\begin{thm}
Let $(A(x),B(x),C(x))$ be the Gosper representation of~$\frac{b(x)}{b(x+m)}$, and~$b(x)p(x,y)$ be a rational solution of \eqref{a(x,y)2}.  Then $b(x)p(x,y)$ must be of the form
\[
b(x)p(x,y)=\frac{B(x-m)p_1(x,y)}{C(x)},
\]
where $p_1(x,y)$ is a polynomial in both $x$ and $y$.
\end{thm}

\pf Assume that
\[
b(x)p(x,y)=\frac{g(x,y)}{q(x)C(x)},
\]
where $g(x,y)\in \mathbb{Q}[x,y]$, $q(x)\in \mathbb{Q}[x]$ is a monic polynomial and
$(q(x),g(x,y))=1$. According to Equation~\eqref{a(x,y)2}, we deduce that
\begin{equation}\label{equation1}
a(x,y)B(x)C(x)q(x)q(x+m)=A(x)g(x+m,y-n)q(x)-B(x)g(x,y)q(x+m).
\end{equation}
It's easy to check that
\[
q(x)\mid g(x,y)B(x)q(x+m).
\]
Since $(q(x),g(x,y))=1$, we obtain
\[
q(x) \mid B(x)q(x+m).
\]
Using this divisibility repeatedly, we can get
\[
q(x) \mid B(x)B(x+m)\cdots B(x+(r-1)m)q(x+rm).
\]
When $r>\max \Disp_x(q(x),q(x))$, we have $(q(x),q(x+rm))=1$, and thus
\[
q(x) \mid B(x)B(x+m)\cdots B(x+(r-1)m).
\]
From Equation~\eqref{equation1}, we also derive that
\[
q(x+m) \mid g(x+m,y-n)A(x)q(x).
\]
By a similar discussion, we arrive at
\[
q(x) \mid A(x-m)A(x-2m)\cdots A(x-rm).
\]
By the definition of Gosper representation, we know that $\gcd(A(x),B(x+hm))=1$ for any $h\in \mathbb{N}$. Thus the only opportunity for $q(x)$ is $q(x)=1$.

When $q(x)=1$, Equation~\eqref{equation1} will be reduced to $$a(x,y)B(x)C(x)=A(x)g(x+m,y-n)-B(x)g(x,y).$$
It's easy to see that
\[
B(x) \mid A(x)g(x+m,y-n),
\]
and hence $B(x) \mid g(x+m,y-n)$.
Setting $g(x,y)=B(x-m)p_1(x,y)$ concludes the proof.
\qed

Substituting~\eqref{Gosper expansion} and~$b(x)p(x,y)=\frac{B(x-m)p_1(x,y)}{C(x)}$ into~\eqref{a(x,y)2}, we obtain
\begin{equation}\label{equation2}
a(x,y)C(x)=A(x)p_1(x+m,y-n)-B(x-m)p_1(x,y).
\end{equation}
Notice that $\deg_y(p_1) <d_0$. Therefore, in order to solve for $p_1(x,y)$, it suffices to find an upper bound on $\deg_x(p_1)$.

From the Gosper representation~\eqref{Gosper expansion}, we see that~$\deg_x(A) = \deg_x(B)$
and their leading coefficients coincide. Now we write
\begin{align*}
& A(x)=\sum_{k=0}^{d_1}a_kx^k, \quad  B(x-m)=\sum_{k=0}^{d_1}b_kx^k,\quad  C(x)=\sum_{k=0}^{d_2}c_kx^k,\\
&a(x,y)=\sum_{k=0}^{d_3}\alpha_k(y)x^k, \quad
 p_1(x,y)=\sum_{k=0}^{d_4}p_k(y)x^k.
\end{align*}
\begin{thm}\label{upbound}
Suppose $A(x)$, $B(x)$ and $C(x)$ are given in $\eqref{Gosper expansion}$, $\deg_y(a)<d_0$ and they have
the above expansions. If $p_1(x,y)$ is a polynomial that satisfies~$\eqref{equation2}$, then
\[
d_4\leq \max \left \{d_2+d_3-d_1+d_0, \,\, \frac{b_{d_1-1}-a_{d_1-1}}{m a_{d_1}}+d_0-1\right \}.
\]
\end{thm}
\pf There are two cases concerning $n$:

\noindent {\it Case~$1.$} $n=0$.

Since the degrees and the leading coefficients of $A(x)$ and $B(x-m)$ coincide, the leading term of the right hand of \eqref{equation2} is canceled. By considering the second leading term, we encounter two cases.

\noindent {\it Case $1a.$} The second leading term is not canceled.  We then have
\[d_4=d_2+d_3-d_1+1.\]

\noindent {\it Case $1b.$} The second leading term is also canceled. We must have
\[a_{d_1}(p_{d_4}(y) d_4  m+p_{d_4-1}(y))+a_{d_1-1} p_{d_4}(y)=a_{d_1} p_{d_4-1}(y)+b_{d_1-1} p_{d_4}(y),\]
which leads to
\[d_{4}=\frac{b_{d_1-1}-a_{d_1-1}}{m a_{d_1}}.\]

\noindent {\it Case~$2.$} $n\neq 0$.

Starting from $i=0$, we consider whether the $(i+1)$th leading term of the right hand of \eqref{equation2} is canceled consequently.

For $i=0$, we have the cases $2a$ and $2b$.

\noindent {\it Case~$2a.$} The leading term is not canceled. We have
\[d_4=d_2+d_3-d_1.\]

\noindent {\it Case~$2b.$} The leading term is canceled.
Then we have
\[p_{d_4}(y-n)=p_{d_4}(y),\]
which implies that $p_{d_4}(y)$ is a constant.

In general, we have the cases~$2a_i$ and $2b_i$.

\noindent {\it Case~$2a_i.$}
The $(i+1)$th leading term is not canceled. Then we have
\[d_4=d_2+d_3-d_1+i.\]

\noindent {\it Case~$2b_i.$}
The $(j+1)$th leading term are all canceled for $j=0,1,2,\ldots,i$.
We claim that $\deg_y (p_{d_4-j}) \le j$ for $j=0,1,2,\ldots,i$.
When $i=0$, the claim holds by the discussion in {\it Case $2b$}.
Suppose we have known that $\deg_y (p_{d_4-j}) \le j$ for $j=0,1,\ldots,i-1$. Now we consider the induction step from $i-1$ to $i$. By the condition that the $(i+1)$th leading term is canceled, we have
\begin{align}
& a_{d_1}\left(p_{d_4-i}(y-n)-p_{d_4-i}(y)\right) \nonumber \\
= & b_{d_1-1} p_{d_4-i+1}(y)+\ldots+b_{d_1-i}p_{d_4} \nonumber \\
  & -a_{d_1}\left(p_{d_4}{d_4\choose i} m^i
¡¡¡¡+p_{d_4-1}(y-n){d_4-1\choose i-1} m^{i-1}+\ldots+ p_{d_4-i+1}(y-n)(d_4-i+1)\right)\nonumber \\
  & -a_{d_1-1}\left(p_{d_4}{d_4\choose i-1} m^{i-1}
    +p_{d_4-1}(y-n){d_4-1\choose i-2} m^{i-2}
    +\ldots+ p_{d_4-i+1}(y-n)\right)\nonumber \\
  & -a_{d_1-2}\left(p_{d_4}{d_4\choose i-2} m^{i-2}
    +p_{d_4-1}(y-n){d_4-1\choose i-3} m^{i-3}
    +\ldots+p_{d_4-i+2}(y-n)\right)\nonumber \\
  & -\ldots-a_{d_1-i} p_{d_4}. \label{eq-deg}
\end{align}
Since $\deg_y(p_{d_4-j})\leq j$ for $j=0,\ldots,i-1$, we know that $\hbox{deg}_y(p_{d_4-i})\leq i$. This proves the claim.

Now consider the above process. Since $\deg_y(p_1(x,y))<d_0$,
there exists an integer $i \le d_0$ such that $\deg_y(p_{d_4-i}) < i$. Without loss of generality, we assume that $i$ is the smallest such integer. If we are in case $2a_i$, we will get an upper bound
\[
d_4=d_2+d_3-d_1+i \le d_2+d_3-d_1+d_0.
\]
Otherwise, we have
\[
\deg_y(p_{d_4-i}(y-n)-p_{d_4-i}(y))\leq i-2,
\]
which means that the coefficient of $y^{i-1}$ on the right hand side of \eqref{eq-deg} is canceled. Thus,
\[b_{d_1-1}-a_{d_1}  (d_4-i+1)  m-a_{d_1-1}=0,\]
leading to the upper bound:
\[d_4=\frac{b_{d_1-1}-a_{d_1-1}}{m a_{d_1}}+i-1 \le \frac{b_{d_1-1}-a_{d_1-1}}{m a_{d_1}}+d_0-1.\]
This completes the proof. \qed

Now we are ready to describe an algorithm for testing the summability of a rational function $f(x, y)$ of the form $f=\frac{a(x,y)}{b(x)d^j(x,y)}$.

\noindent \emph{Algorithm SumTest:}

\noindent \emph{Input:} A rational function~$f=\frac{a(x,y)}{b(x)d^j(x,y)}$.\\
\emph{Output:} True, if~$f$ is~$(\sigma_x, \sigma_y)$-summable and return~$p$ such that Equation~\eqref{EQ:denom} holds;
False, otherwise.

1. Compute $\Disp(d(x,y),d(x,y))$ by algorithm~\emph{DispSet}. If it is the set $\{(0,0)\}$, then return False. Otherwise,
let $(m,n)$ be the element in the dispersion set such that $m$ is the minimum positive integer.

2. Compute the Gosper representation~${(A(x),B(x),C(x))}$ of $\frac{b(x)}{b(x+m)}$.

3. Set
\[
d_0=\deg_y(d(x,y)),\ d_1=\deg_x(A(x)),\ d_2=\deg_x(C(x)),\ \mbox{and}\ d_3=\deg_x(a(x,y)).
\]
Suppose
\begin{align*}
& a_{d_1} =\hbox{coeff}(A(x),x,d_1), \\
& a_{d_1-1} =\hbox{coeff}(A(x),x,d_1-1), \\ & b_{d_1-1}=\hbox{coeff}(B(x-m),x,d_1-1),
\end{align*}
where $\hbox{coeff}(f(x),x,m)$ denotes the coefficient of $x^m$ in the expansion of $f(x)$.

4. Let $d_4 = \hbox{max}\{d_2+d_3-d_1+d_0,\, \,
                       \left\lfloor\frac{ b_{d_{1}-1}-a_{d_{1}-1}}{ma_{d_1}}\right\rfloor+d_0-1\}$.
  Set
  \[
  p_1(x,y)=\sum_{i=0}^{d_4} \sum_{j=0}^{d_0-1} c_{i,j} x^i y^j,
  \]
  and plug it into
\[
a(x,y)C(x) = A(x)p_1(x+m,y- n)- B(x-m)p_1(x,y).
\]
By comparing all the coefficients of $x$ and $y$ on both sides, we
determine whether the above equation has a solution $p_1(x,y)$.
If it has no solution, then return False. Otherwise, return a solution
$p_1(x,y)$ and $p(x,y)=\frac{B(x-m)p_1(x,y)}{b(x)C(x)}$.

Finally, we give some examples to illustrate how to
use our criterion for deciding the summability of some rational functions.
\begin{ex}
Let
\[
f(x,y)=-\frac{(x+y+4)}{(x^2+2x+2xy-1+2y+y^2)(x^2+2xy+y^2-2)}.
\]
Denote
\[
d(x,y) = x^2+2xy+y^2-2 = (x+y)^2-2.
\]
By computing the dispersion set, we find that
\[
x^2+2x+2xy-1+2y+y^2 = d(x+1,y).
\]
By partial fraction decomposition, we derive that
$$f(x,y)=\sum_{l=0}^1\frac{a_l(x,y)}{d(x+l,y)},$$
where
\[
a_0(x,y)=-x-y,\quad a_1(x,y)=x+y+2.
\]
Using~$(\sigma_x, \sigma_y)$-reduction, we can write $f(x,y)$ as
\begin{equation}\label{equation3}
f(x,y)=\Delta_x(g_1) +r(x,y),
\end{equation}
where
\[
g_1(x,y)=\frac{x+y+1}{(x+y)^2-2} \quad \mbox{and} \quad r(x,y)=\frac{1}{(x+y)^2-2}.
\]

It is easy to see that $\sigma_x d(x,y)=\sigma_y d(x,y)$. What left now is to check whether there exists a  polynomial $p(x,y)$ such that
\begin{equation}\label{equation4}
1=\sigma_x\sigma_y^{-1}p(x,y)-p(x,y).
\end{equation}
The $x$-degree bound and $y$-degree bound of $p$ are $1$ and $2$ respectively. By the method of undetermined coefficients, we find a solution $p(x,y)=-y-1$. From the proof of Lemma~\ref{lemma5}, we  find out
\[
r(x,y)=\sigma_xg_2(x,y)-g_2(x,y)+\sigma_yh(x,y)-h(x,y),
\]
where
$$g_2(x,y)=\frac{-y-1}{(x+y)^2-2},\ h(x,y)=\frac{y}{(x+y)^2-2}.$$
Substituting into \eqref{equation3}, we finally derive that
\[
f(x,y)=\sigma_xg(x,y)-g(x,y)+\sigma_yh(x,y)-h(x,y),
\]
where
\[
g(x,y)=\frac{x}{(x+y)^2-2}, \quad \mbox{and} \quad h(x,y)=\frac{y}{(x+y)^2-2}.
\]
\end{ex}

\begin{ex}
Let
\[
f(x,y)=\frac{x^2+x^2y+y^2+1}{(x^2+y^2)(x^3+2xy+xy^2+y^3)}.
\]
We can decompose it into
$$f(x,y)=\frac{x^5+x^4-2x+x^2-2x^3+x^4y-xy-y^2}{(x^3+2xy+xy^2+y^3)x^3(x-2)}+
\frac{-x^4+y}{(x^2+y^2)x^3(x-2)}.$$
Note that
\[
\sigma_x^m(x^2+y^2)\neq\sigma_y^n(x^2+y^2),\quad\hbox{for any }(m,n)\neq (0,0).
\]
Then Theorem~\ref{theorem2} implies that $\frac{-x^4+y}{(x^2+y^2)x^3(x-2)}$ is not $(\sigma_x,\sigma_y)$\hbox{-}summable, which leads to the result that $f(x,y)$ is not $(\sigma_x,\sigma_y)$\hbox{-}summable in $\mathbb{Q}(x,y)$.
\end{ex}

\noindent \textbf{Acknowledgments.} This work was supported by the 973 Project, the PCSIRT Project of the Ministry of Education and the National Science Foundation of China.

\def\polhk#1{\setbox0=\hbox{#1}{\ooalign{\hidewidth
  \lower1.5ex\hbox{`}\hidewidth\crcr\unhbox0}}}

{\section*{Appendix}\label{SECT:appendix}
In Theorem~\ref{theorem2}, we provide a criterion for the summability of rational functions of the
form~$\frac{a(x,y)}{d^j(x,y)}$, where~$a\in \bQ(x)[y]$ and~$d \in \bQ[x,y]$ is an irreducible polynomial.
In this appendix, we present the proof of this criterion.

Firstly, we give the following lemma which proves the sufficiency of Theorem~\ref{theorem2}.
\begin{lem}\label{lemma5}
Let $f \in \mathbb{Q}(x,y)$ be of the form $f =\frac{a(x,y)}{d^j(x,y)}$,
where~$d\in \mathbb{Q}[x,y]$ is an irreducible polynomial, $a\in \mathbb{Q}(x)[y]$, and $\deg_y(a)<\deg_y(d)$. Suppose that there exist integers $t,l$ with $t>0$ and a polynomial $p(x,y)\in \mathbb{Q}(x)[y]$ such that
\[\sigma_x^{t}d(x,y)=\sigma_y^{\ell}d(x,y),\]
and
\[a=\sigma_x^{t}\sigma_y^{-\ell}p(x,y)-p(x,y).\]
Then $f $ is $(\sigma_x,\sigma_y)$\hbox{-}summable in $\mathbb{Q}(x,y)$.
\end{lem}
\pf Let $g =\sum_{k=0}^{t-1}\frac{\sigma_x^k (p)}{\sigma_x^k (d^j)}$, then
\[
   \frac{a}{d^j}-(\sigma_x g - g )  =  \frac{a}{d^j}-\frac{\sigma_x^{t}p}{\sigma_x^{t}d^j}+\frac{p}{d^j}
  =  \frac{a+p}{d^j}
    -\frac{\sigma_x^{t}p}{\sigma_y^{\ell}d^j}
 =  -\sigma_y^{\ell}\left(\frac{\sigma_x^{t}\sigma_y^{-\ell}p}{d^j}\right) +\frac{\sigma_x^{t}\sigma_y^{-\ell}p}{d^j}. \tag*{\qed}
\]

The rest part of the appendix is devoted to proving the necessity of Theorem~\ref{theorem2}.

Analogue to the discrete residue given by Chen and Singer \cite{ChenSinger2014}, we introduce the concept of polynomial residue. Let $\mathbb{K}$ be a field and $ f \in \mathbb{K}(x)$. By partial fraction decomposition, $f$ can be  written as
\begin{equation}\label{EQ:univariate}
f =p(x)+\sum_{i=1}^m\sum_{j=1}^{n_i}\sum_{\ell=0}^{k_{i,j}}
     \frac{a_{i,j,\ell}(x)}{\sigma_x^{\ell}d_i^j(x)},
\end{equation}
where $p(x)\in \mathbb{K}[x]$, $m,n_i,k_{i,j}\in \mathbb{N}$,
$\deg_x(a_{i,j,\ell})<\deg_x(d_i)$,  and $d_i(x)\, (i=1,\ldots, m)$ are irreducible polynomials that in distinct $\sigma_x$-orbits. The summation
\[
\sum_{\ell=0}^{k_{i,j}}\sigma_x^{-\ell}(a_{i,j,\ell})
\]
is called the \emph{polynomial residue} of
$f$ at the $\sigma_x$-orbit of $d_i(x)$ of multiplicity $j$, denoted by $\res_x(f(x),d_i(x),j)$.

It is easy to check that the summability of rational functions in $\mathbb{K}(x)$ can be given via polynomial residue. The proof is similar to the case of discrete residue~\cite{ChenSinger2012, ChenSinger2014} and is omited.

\begin{prop}\label{univariate}
Let $f(x)\in \mathbb{K}(x)$ be of the form \eqref{EQ:univariate}.
Then $f(x)$ is $\sigma_x$-summable in $\mathbb{K}(x)$ if and only if the
polynomial residue~$\res_x(f(x),d_i(x),j)$ is zero for any polynomial $d_i(x)$ and any multiplicity~$j$.
\end{prop}

Now we are ready to prove the necessity of Theorem~\ref{theorem2}.

Suppose that $f=a/d^j$ is $(\sigma_x,\sigma_y)$-summable and assume that
\begin{equation}\label{summable}
f =\sigma_x g - g +\sigma_y h -h,
\end{equation}
where $g,h \in \mathbb{Q}(x,y)$.
As a univarite analogue to Lemma~\ref{LM:xy-reduce},
we can decompose~$g$ into the form
\[
g =\sigma_y g_1 - g_1 + g_2 +\frac{\lambda_1}{\sigma_x^{\mu_1}d^j}
           +\cdots+\frac{\lambda_s}{\sigma_x^{\mu_s}d^j},
\]
where $g_1 , g_2 \in \mathbb{Q}(x,y)$ with $g_2$ containing no term of the form
$\frac{\lambda}{\sigma_x^{u}d^j}$
in its partial fraction decomposition with respect to $y$, $\mu_{\ell}\in \mathbb{Z}$, $\lambda_\ell \in \mathbb{Q}(x)[y]$,
and $\sigma_x^{\mu_{\ell}}d\, (\ell=1,\ldots, s)$ are irreducible polynomials in distinct $\sigma_y$\hbox{-}orbits.

\noindent
\textbf{Claim 1.} Let
\[
\Lambda:=\{\sigma_x^{\mu_1}d,\ldots,\sigma_x^{\mu_s}d,
\sigma_x^{\mu_1+1}d,\ldots, \sigma_x^{\mu_s+1}d\}.
\]
Then
\begin{itemize}
\item[(a)] At least one element of $\Lambda$ is in the same $\sigma_y$\hbox{-}orbit as $d$.
\item[(b)] For each element $\eta\in \Lambda$, there is one element of $\Lambda\backslash\{\eta\}\bigcup\{d\}$ that is in
the same $\sigma_y$\hbox{-}orbit as $\eta$.
\end{itemize}

\noindent {\it Proof of Claim 1.} (a) Suppose there is no element of $\Lambda$ that is in the same $\sigma_y$\hbox{-}orbit as $d$. Since $f=a/d^j$, we have $\res_y(f,d,j)=a \neq0$. While by~\eqref{summable} and Proposition~\ref{univariate}, we deduce that
\[
\res_y(f,d,j) = \res_y(\sigma_x g - g, d,j)=0,
\]
which is a contradiction.

(b) The assertion follows from the same argument when considering the polynomial residues
of $\eta$ on both sides of~\eqref{summable}. \qed

Claim 1 implies that either $d \sim_y \sigma_x^{\mu_1'}d\hbox{ or }\ d \sim_y \sigma_x^{\mu_1'+1} d$ for some $\mu_1'\in \{\mu_1,\ldots,\mu_s\}$. We will only consider the first case. The second case can be treated similarly.

\noindent
\textbf{Claim 2.} Assume $d \sim_y \sigma_x^{\mu_1'} d$. We have the following assertions.

(a) Suppose $k\geq 2$ be an integer such that $\sigma_x^ld\nsim_yd$ for $1\leq l\leq k-1$. Then there exist $\mu_1',\ldots,\mu_k'\in\{\mu_1,\ldots,\mu_s\}$ such that
\[
 \sigma_x^{\mu_1'+1}d \sim_y \sigma_x^{\mu_2'}d,\quad
 \sigma_x^{\mu_2'+1}d \sim_y \sigma_x^{\mu_3'}d,\quad
 \ldots, \quad
 \sigma_x^{\mu_{k-1}'+1}d \sim_y \sigma_x^{\mu_k'}d,
\]
and
\[\sigma_x^{k-1}d \sim_y \sigma_x^{\mu_k'}d.\]

(b) There exists a positive integer $t\leq s$ such that $\sigma_x^{t}d\sim_y d$.

\noindent {\it Proof of Claim 2.} (a) From Claim~$1$(b), we derive that
$\sigma_x^{\mu_1'+1}d$ is $\sigma_y$\hbox{-}equivalent to an element of $\Lambda\backslash\{ \sigma_x^{\mu_1'+1} d\}\bigcup\{d\}$.
If $\sigma_x^{\mu_1'+1}d \sim_y d$, then
$\sigma_x^{\mu_1'+1} d \sim_y \sigma_x^{\mu_1'}d$ and thus
$\sigma_x d\sim_yd$, which contradicts to the hypothese on $k$.
If $\sigma_x^{\mu_1'+1}d \sim_y \sigma_x^{\mu_l'+1}d$, then $\sigma_x^{\mu_1'}d \sim_y \sigma_x^{\mu_l'}d$ for some $l$, which contradicts to the assumption that $\sigma_x^{\mu_\ell}$ are in distinct $\sigma_y$-orbits. Therefore we are left with the only possibility that
$\sigma_x^{\mu_1'+1} d\sim_y \sigma_x^{\mu_2'} d$ for some $\mu_2' \in \{\mu_1,\ldots,\mu_s\}\setminus \{\mu_1'\}$. Continue this process, we will find $\mu_3', \ldots, \mu_k'$ such that
\[
 \sigma_x^{\mu_2'+1}d \sim_y \sigma_x^{\mu_3'}d,\quad
 \ldots, \quad
 \sigma_x^{\mu_{k-1}'+1}d \sim_y \sigma_x^{\mu_k'}d.
\]
Finally, we have
\[
\sigma_x^{\mu_k'} d \sim_y \sigma_x^{\mu_1'+k-1} d \sim_y \sigma_x^{k-1} d.
\]

(b) If such $t$ does not exist, then one could find $\{\mu_1',\ldots,\mu_{s+1}'\}$ satisfying the constraints in (a). Thus, it holds that $\mu_r'=\mu_t'$ for some $r>t$. Hence
$\sigma_x^{\mu_1'+r}d \sim_y \sigma_x^{\mu_1'+t}d$, which leads to
$\sigma_x^{r-t}d\sim_yd$, a contradiction. \qed

Suppose $t$ is the smallest integer such that $\sigma_x^t d \sim_y d$. Then taking $k=t$ in Claim 2(a), we derive that
there exist $\mu_1',\ldots,\mu_{t}'\in\{\mu_1,\ldots,\mu_s\}$ such that
\[
 \sigma_x^{\mu_1'+1} d \sim_y \sigma_x^{\mu_2'}d, \quad
 \sigma_x^{\mu_2'+1} d \sim_y \sigma_x^{\mu_3'}d, \quad
 \ldots, \quad
 \sigma_x^{\mu_{t-1}'+1}d \sim_y \sigma_x^{\mu_{t}'}d,
\]
and
\[
\sigma_x^{\mu'_t+1}d \sim_y \sigma_x^t d \sim_y d.
\]
Recall that $\sigma_x^{\mu_1'} d \sim_y d$. By the definition of $\sim_y$, there exist integers $s_0,s_1,\ldots,s_t$ such that
\[
\sigma_x^{\mu_k'+1} d=\sigma_x^{\mu_{k+1}'}\sigma_y^{s_{k+1}}d, \ 1\leq k \leq t-1, \quad \sigma_x^{\mu_{t}'+1}d=\sigma_y^{s_1}d, \quad \mbox{and} \quad \sigma_x^{\mu_1'} d=\sigma_y^{s_0} d.
\]
Hence,
\[
\sigma_y^{s_0} d = \sigma_x^{\mu_1'}d = \sigma_x^{\mu'_2-1}\sigma_y^{s_2} d = \sigma_x^{\mu'_3-2}\sigma_y^{s_2+s_3} d = \cdots =
\sigma_x^{\mu'_t-t+1}\sigma_y^{s_2+s_3+\cdots+s_t} d = \sigma_y^{s_1+\cdots+s_t}\sigma_x^{-t} d.
\]
Setting $\ell=s_1+\cdots + s_t - s_0$, we then have
$\sigma_x^t d = \sigma_y^\ell d$.

Now we compare the polynomial residues on both sides of \eqref{summable}.
\begin{table}[ht]
\renewcommand{\arraystretch}{1.0}
\begin{center}
\begin{tabular}{|c|c|c}
\hline \rule{0pt}{15pt}
 $\sigma_y$-orbit& Comparison of two sides of \eqref{summable} \\[3pt]
\hline \rule{0pt}{15pt}
$d,\ \sigma_x^{\mu_{t}'+1}d$ & $
a=\sigma_x \sigma_y^{-s_0}\lambda_{t}'-\sigma_y^{-s_1} \lambda_1'$ \\[3pt]
\hline \rule{0pt}{15pt}
$\sigma_x^{\mu_{t-1}'+1}d,\ \sigma_x^{\mu_{t}'}d$ & $0=\sigma_x\sigma_y^{-s_{t}}\lambda_{t-1}'-\lambda_{t}'$  \\[3pt]
\hline \rule{0pt}{15pt}
$\sigma_x^{\mu_{t-2}'+1}d,\ \sigma_x^{\mu_{t-1}'}d$ &
$0=\sigma_x\sigma_y^{-s_{t-1}}\lambda_{t-2}'-\lambda_{t-1}'$ \\[3pt]
\hline
$\vdots$ & $\vdots$ \\
\hline \rule{0pt}{15pt}
$\sigma_x^{\mu_2'+1}d,\  \sigma_x^{\mu_3'}d$ &
$0=\sigma_x \sigma_y^{-s_3}\lambda_2'-\lambda_3'$ \\[3pt]
\hline \rule{0pt}{15pt}
$\sigma_x^{\mu_1'+1}d,\ \sigma_x^{\mu_2'}d$ &
$0=\sigma_x\sigma_y^{-s_2}\lambda_1'-\lambda_2'$ \\[3pt]
\hline
\end{tabular}
\end{center}
\caption{Orbits and their corresponding polynomial residues. \label{table}}
\end{table}
We list the residues in Table~\ref{table}, where the first column consists of the
$\sigma_y$\hbox{-}orbits of elements in $\Lambda$ and the second column consists of the equations obtained by equating the corresponding polynomial residues on both sides of \eqref{summable}. By investigating the equations in Table~\ref{table} from bottom to top, we find that
\[
a=\sigma_x^{t}\sigma_y^{-\ell}p-p,
\]
where $p=\sigma_y^{-s_1}\lambda_1'(x,y)$. Since $\deg_y \lambda_1' < \deg_y d$, we have $\deg_y p < \deg_y d$. This completes the proof of Theorem~\ref{theorem2}.\qed

\end{document}